\documentclass[useAMS,usenatbib]{mn2e}
\usepackage[pdftex]{graphicx}
\def\apj{ApJ}
\def\apjl{ApJL}
\def\aap{A\&A}
\def\mnras{MNRAS}
\def\nat{Nature}
\def\prd{Phys.~Rev.~D}

\title[Thermal emission from the exoplanet WASP-19b]{Ground-based detection of thermal emission from the exoplanet WASP-19b
\thanks{
Based on observations collected with the HAWK-I instrument at the VLT/UT4 Yepun telescope at the European Organisation for Astronomical Research in the Southern Hemisphere, Chile (Programme: 282.C-5019)}
}
\author[N. P. Gibson et al.]{
N. P. Gibson$^{1,2,3}$\thanks{E-mail: Neale.Gibson@astro.ox.ac.uk},
S. Aigrain$^{1,2}$,
D. L. Pollacco$^{3}$,
S. C. C. Barros$^{3}$,
L. Hebb$^{4}$, \newauthor
M. Hrudkov\'a$^{5}$,
E. K. Simpson$^{3}$,
I. Skillen$^{6}$ and
R. West$^{7}$
\\
$^{1}$Department of Physics, University of Oxford, Denys Wilkinson Building, Keble Road, Oxford OX1 3RH, UK\\
$^{2}$School of Physics, University of Exeter, Exeter, EX4 4QL, UK\\
$^{3}$Astrophysics Research Centre, School of Mathematics \&\ Physics, Queen's University, University Road, Belfast, BT7 1NN, UK\\
$^{4}$Physics \& Astronomy Department, Vanderbilt University, 6301 Stevenson Center, Nashville, TN 37235, USA\\
$^{5}$Th\"{u}ringer Landessternwarte Tautenburg, Sternwarte 5, D - 07778 Tautenburg, Germany \\
$^{6}$Isaac Newton Group of Telescopes, Apartado de Correos 321, E-38700 Santa Cruz de la Palma, Tenerife, Spain\\
$^{7}$Department of Physics and Astronomy, University of Leicester, Leicester, LE1 7RH, UK
}

\begin{document}

\date{Accepted 1988 December 15. Received 1988 December 14; in original form 1988 October 11}

\pagerange{\pageref{firstpage}--\pageref{lastpage}} \pubyear{2002}

\maketitle

\label{firstpage}

\begin{abstract}
We present an occultation of the newly discovered hot Jupiter system WASP-19, observed with the HAWK-I instrument on the VLT, in order to measure thermal emission from the planet's dayside at $\sim$2$\umu$m. The light curve was analysed using a Markov-Chain Monte-Carlo method to find the eclipse depth and the central transit time. The transit depth was found to be 0.366$\pm$0.072\%, corresponding to a brightness temperature of 2540$\pm$180\,K. This is significantly higher than the calculated (zero-albedo) equilibrium temperature, and indicates that the planet shows poor redistribution of heat to the night side, consistent with models of highly irradiated planets. Further observations are needed to confirm the existence of a temperature inversion, and possibly molecular emission lines. The central eclipse time was found to be consistent with a circular orbit.

\end{abstract}


\section{Introduction}

The vast majority of exoplanets are discovered due to effects they have on the observed orbits and brightness of their host stars. Transiting planets are detected when the planet passes in front of its host star, causing the amount of light we detect to decrease periodically. From the shape of the transit light curve, coupled with radial velocity measurements, we can determine the radius and mass of the planet, and infer its composition.

During occultation, when the planet passes behind its host star, we have the opportunity to directly measure the flux emitted by the planet. This allows detection of thermal emission from the planet's dayside and reflected starlight, and in the near-infrared (where the flux from the planet is dominated by thermal emission), yields a measurement of the planet's temperature. Occultations of transiting planets were first detected from space using the Spitzer space telescope \citep{charbonneau,deming}, which has continued to measure thermal emission at wavelengths $\ge3.6\,\umu$m. Ground based infrared photometry should provide complementary information, providing measurements of thermal emission at shorter wavelengths. This is the region where the flux from a typical hot Jupiter is expected to reach its peak, and also where molecular absorption (or emission) bands in the atmospheres play an important role \citep[see e.g.][]{barman_2005,marley_2007,fortney_2008}. 

However, ground based infrared photometry of occultations has proven extremely difficult to date, due to the very high signal-to-noise requirements and the relatively poor stability of IR detectors in comparison to optical CCDs. Following some tentative detections \citep[e.g.][]{snellen_covino}, the first ground based detections of thermal emission were recently published by \citet[][TrES-3b in K-band]{demooij_snellen}, \citet[][OGLE-TR-56b in z' band]{sing_lopez} and \citet[][CoRoT-1b at $\sim2\umu$m]{gillon_2009}.

Here, we present observations of an occultation of the newly discovered transiting exoplanet system WASP-19 \citep{hebb_2010}. WASP-19 is a G-dwarf star with an effective temperature of $5500\pm100$\,K, with a $1.15\pm0.08$ Jupiter-mass planet in a very short period orbit of $0.7888399\pm0.0000008$ days. The planet is therefore expected to be extremely hot due to the proximity to its host, and coupled with its large radius of $1.31\pm0.06$ Jupiter-radii is expected to produce an occultation in the K-band $\simeq0.2-0.4\%$. It was observed using the HAWK-I (High Acuity Wide-field K-band Imager) instrument \citep{pirard,casali,kissler}, a wide-field, near-infrared camera mounted on UT4 of the VLT, which was recently used by \citet{gillon_2009} to detect the occultation of CoRoT-1b. Simultaneously with the original submission of the present manuscript, \cite{anderson_2010} announced an independent detection of the secondary eclipse of WASP-19b in the H-band, also with HAWK-I, finding an eclipse depth of {$0.259^{+0.046}_{-0.044}\%$}. In Section~\ref{sect:observations}, we describe the observations and data reduction. Sections~\ref{sect:modelling}~and~\ref{sect:results} present our results and analysis, and finally in Section~\ref{sect:summary} we summarise and discuss our findings.

\section[]{HAWK-I observations and data reduction}
\label{sect:observations}

\begin{figure}
\includegraphics[width=84mm]{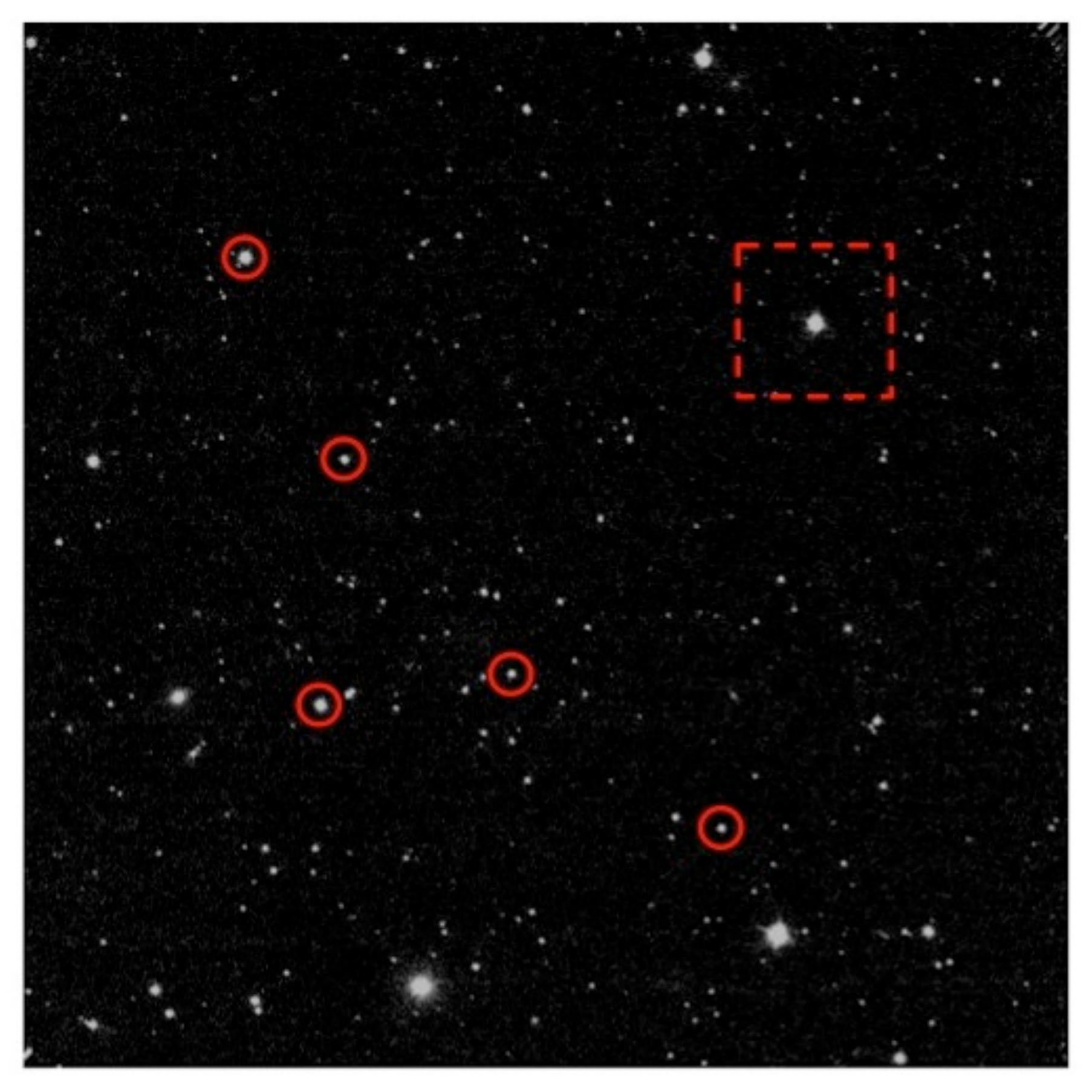}
\caption{
HAWK-I image of quadrant 2 showing the position of WASP-19 (box) and the five comparison stars used for relative photometry (circles). The box around WASP-19 shows the size of the dithering box used by the AutoJitter template.
}
\label{fig:field_of_view}
\end{figure}

An occultation of the exoplanet system WASP-19 was observed using HAWK-I. The observations were carried out in service mode on the night of March 30 2009, and lasted for $\sim$4.6 hours, starting $\sim$1.5 hours before the expected ingress, and ending $\sim$1.6 hours after expected egress (assuming the planet is in a circular orbit). HAWK-I consists of four Hawaii 2RG 2048$\times$2048 pixel detectors, arranged in a mosaic giving a total field of view of 7.5 arcminutes squared, with a pixel scale of 0.106 arcseconds. However, in order to obtain very precise relative photometry needed to detect a secondary transit, we found \citep[similar to][]{gillon_2009} that only reference stars that fell on the same chip (quadrant 2) as WASP-19 could be used, reducing the effective field of view. An image of the HAWK-I field (quadrant 2 only) is shown in Fig.~\ref{fig:field_of_view}.

Observations were taken using the NB2090 filter, a narrow filter centred at 2.095\,$\umu$m, in the region where the thermal emission of hot Jupiters ($\sim1000-1800\,$K) is expected to peak. Using a narrow filter should also help reduce differential extinction. This still provides high enough signal-to-noise to detect the occultation assuming observations are photon limited, in which case we would obtain sub-millimagnitude photometry per 10 second exposure\footnote{According to the VLT/HAWK-I exposure time calculator http://www.eso.org/observing/etc/ with K = 10.5.}. However, from experience we know observations of this type are limited by systematic noise. The telescope was defocussed slightly to keep the stars well within the linearity range of the detectors, and to spread the stellar profile over more pixels.

\begin{figure}
\includegraphics[width=84mm]{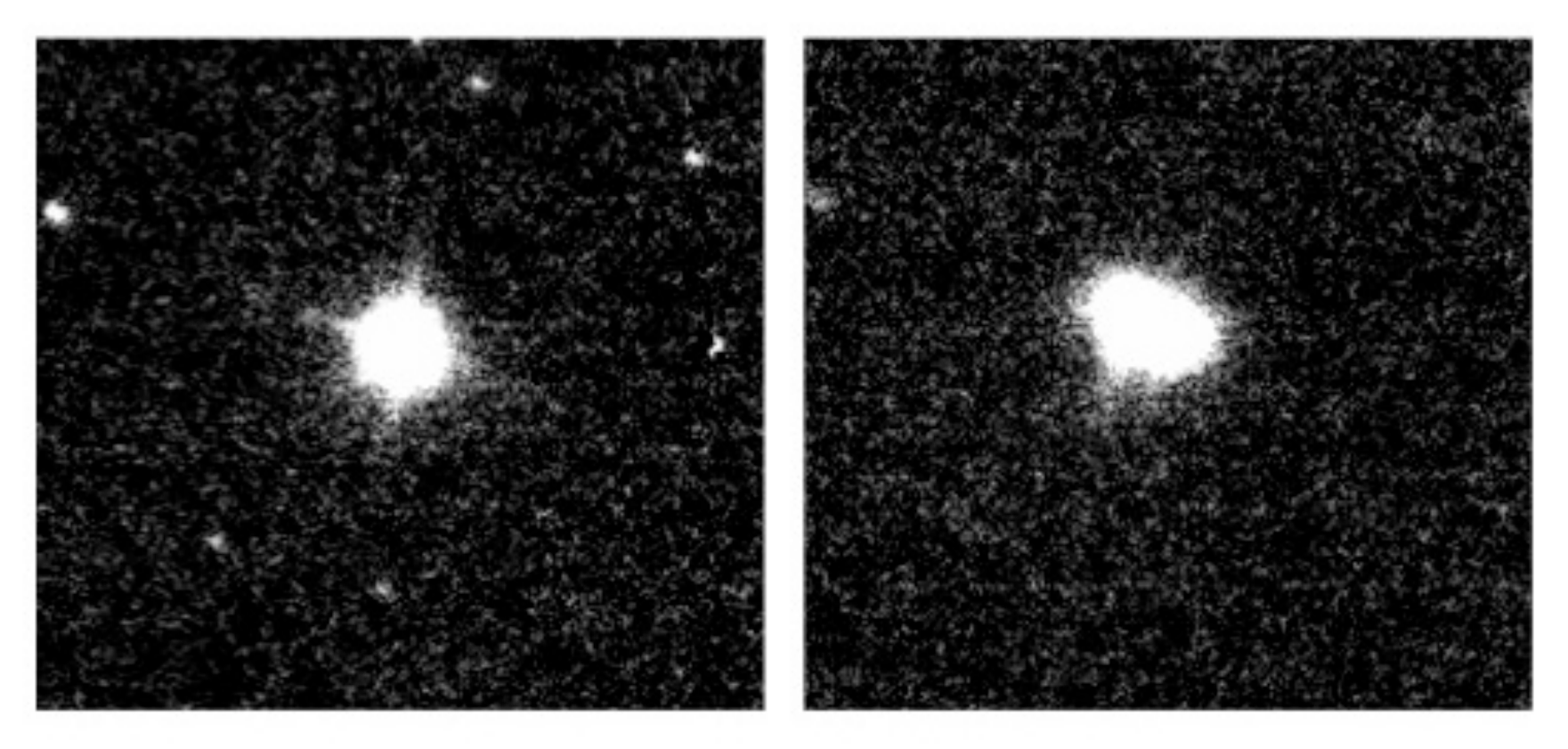}
\caption{
Close-ups of WASP-19 in the HAWK-I images showing how the PSFs elongated during the course of the observations. The left image shows a typical PSF during the first half of the night with a round shape. The right image shows a typical PSF towards the end of the night, with an elongated shape.
}
\label{fig:focus}
\end{figure}

A total of 185 images were acquired, each consisting of $6\times10$ second exposures. A dithering pattern was created at random within 30 arsceconds of the first image using the AutoJitter template. A basic calibration including bias subtraction and flat-fielding was performed on each image using the EsoRex HAWK-I pipeline. Sky-subtraction was also performed using EsoRex. For each image, a sky background frame was constructed from a running median of the object frame and the nine previous and nine subsequent frames. This process involved two passes, the second of which masked objects from the background calculation. The stars had a typical FWHM of 0.5--0.6 arcseconds, but around half way through the observations the PSFs of the stars deteriorated and began to elongate, as shown in Fig.~\ref{fig:focus}, probably related to the defocussing of the telescope. The occultation signal could still be recovered from the photometry, providing large enough apertures were used. However, this is likely responsible for the large gradient seen in the raw light curve.

\begin{figure}
\includegraphics[width=84mm]{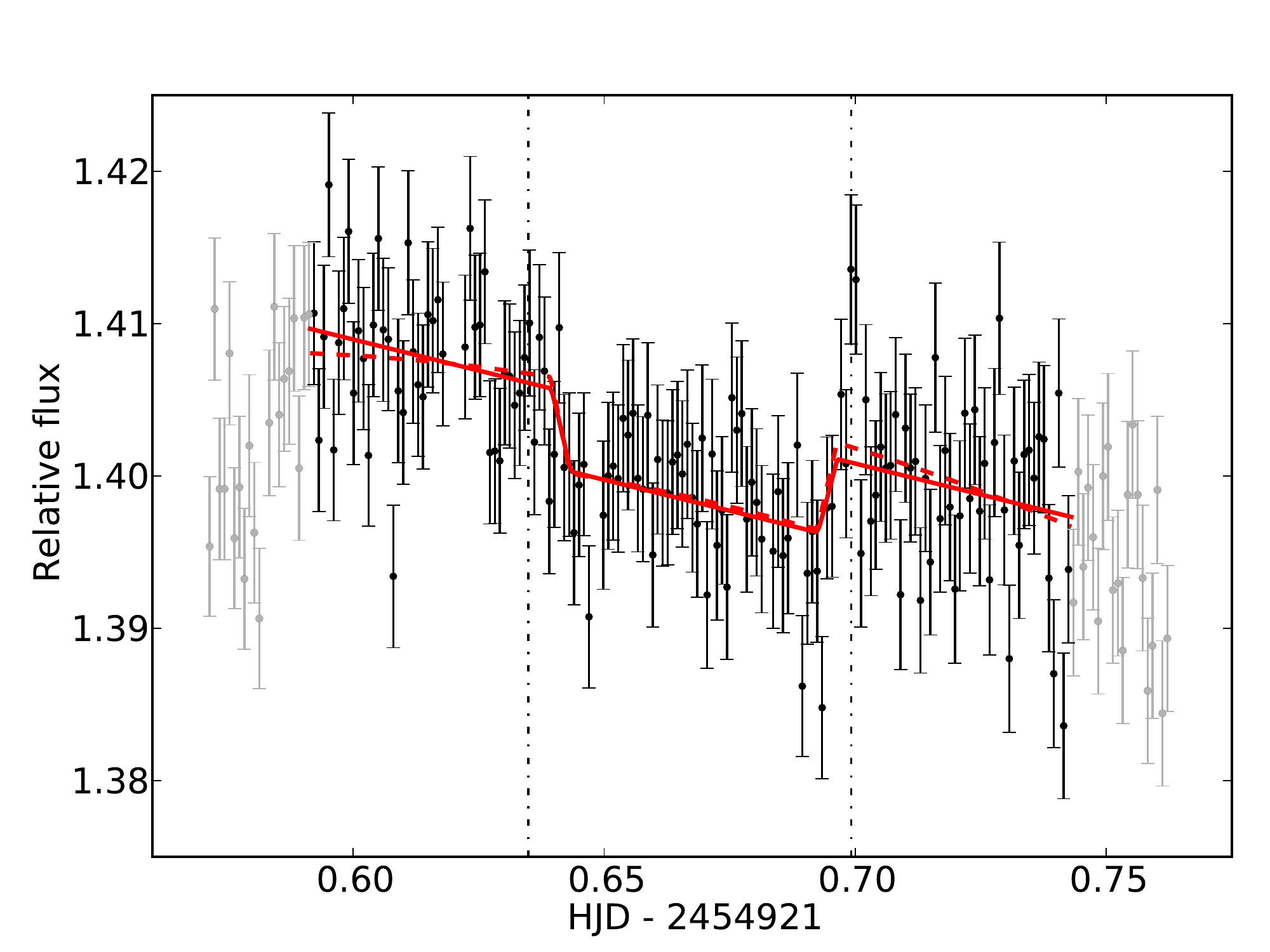}
\caption{
Raw VLT/HAWK-I light curve of the secondary transit of WASP-19. The dashed-dotted lines show the expected start and end of transit, assuming the planet is in a circular orbit. A transit is clearly visible at the expected time and duration. The solid and dashed lines show the best-fit models using the linear and quadratic normalisation functions, respectively. The first and last 20 images were not used in the fitting process, and the corresponding data points are shown in grey. The error bars shown were re-scaled so that the recuded $\chi^2$ of the best-fit model (using the linear normalisation) is equal to one.}
\label{fig:lcv_raw}
\end{figure}

Aperture photometry was performed on WASP-19 and five comparison stars using Pyraf\footnote{Pyraf is a product of the Space Telescope Science Institute, which is operated by AURA for NASA.} and the DAOPHOT package. The positions of WASP-19 and the five comparison stars used are marked in Fig.~\ref{fig:field_of_view}. Several brighter comparison stars were ruled out either because they were over the linearity limit of the chip, or because they fell too near the edge of the field on some images due to the dithering. An aperture of 12 pixel radius was used for each star, and the remaining sky was estimated using an annulus of width 5 pixels beginning at 20 pixels from the stars' centres. The flux of WASP-19 was then divided by the sum of the flux from the five comparison stars to obtain the raw light curve, shown in Fig.~\ref{fig:lcv_raw}. Initial estimates of the photometric errors were calculated using the aperture electron flux, sky and read noise.

Clearly, a trend can be seen in the light curve, likely caused by the PSFs elongating during the observations. This was removed by normalising the light curve by fitting a linear or quadratic function of time to the out-of-transit data, to set the out-of-transit flux equal to 1. The best-fit models using the linear and quadratic functions are shown in Fig.~\ref{fig:lcv_raw}, although the linear function is preferred and used to determine the light curve parameters (see Sect.~\ref{sect:modelling}). The normalisation of the light curves for both functions was strongly affected by outlying points at the start and end of the observations. As the sky background was estimated from 9 frames at either side of each image, the sky subtraction was not estimated accurately. We decided to remove the first and last 20 data points ($\sim30$ minutes) from the light curve analysis (shown in grey in Fig.~\ref{fig:lcv_raw}), which still left substantial out-of-transit data to constrain the normalisation function.

\section[]{Light curve modelling and analysis}
\label{sect:modelling}

\begin{figure}
\includegraphics[width=84mm]{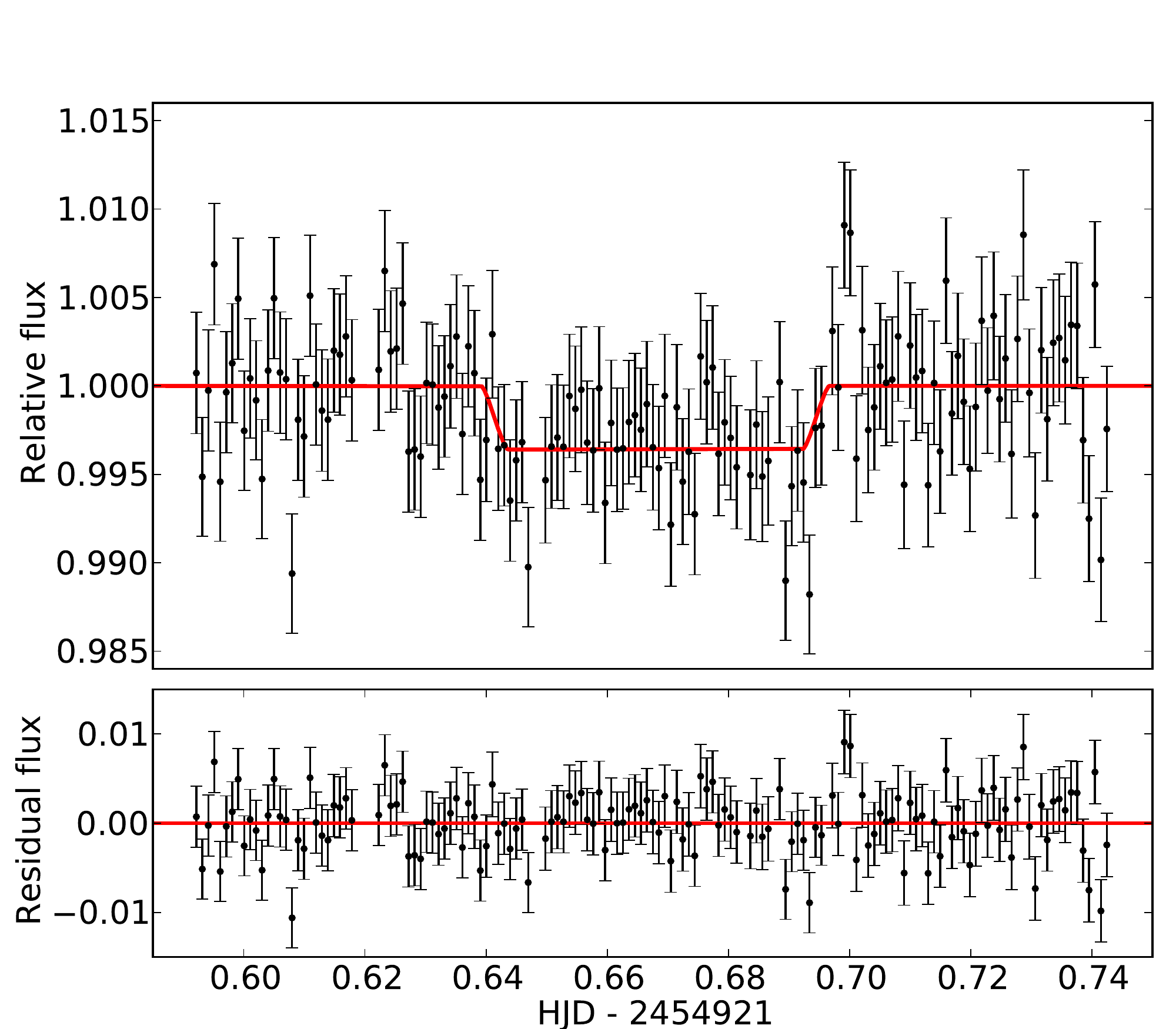}
\caption{
Top:VLT/HAWK-I light curve of the secondary transit of WASP-19 normalised with a linear function of time, with the best-fit model over-plotted and with the first and last 20 images removed. Bottom: Residuals from the best-fit model.}
\label{fig:lcv_norm}
\end{figure}

The HAWK-I light curve was fitted using an MCMC (Markov-Chain Monte-Carlo) routine, which is a method used to explore the multi-dimensional parameter space of a model fit efficiently, allowing a determination of the joint posterior probability distribution for the parameters \citep[see e.g.,][]{tegmark_2004, holman_winn_2006, cameron_2007, winn_holman_2008}. Our implementation of MCMC uses the $\chi^2$ fitting statistic for a model light curve given by
\[
\chi^2=\sum_{j=1}^{N}\frac{(f_{{\rm obs},j} - f_{{\rm calc},j})^2}{\sigma^2_j},  
\]
where  $f_{{\rm obs},j}$ is the flux observed at time $j$, $\sigma_j$ is the corresponding uncertainty and  $f_{{\rm calc},j}$ is the flux calculated from the model for time $j$.

The model flux $f_{{\rm calc}}$ was constructed using Kepler's laws to determine the normalised separation of the planet and star centres as a function of time assuming a circular orbit. The analytic equations of \citet{mandel_agol_2002} were then used to calculate the flux from the normalised separation assuming no limb darkening on the planet's surface (as this will only affect the ingress and egress which the HAWK-I data is unable to constrain). The system parameters were held fixed at the values determined by \citet{hebb_2010} to determine the shape of the transit. The depth $D$ and width $W$ of the transit could then be varied by simply scaling the transit shape in both flux and time axes.

A long chain of model parameter sets is created by adding small Gaussian perturbations to the previous accepted set. At each step we apply the Metropolis-Hastings rule to decide whether to accept the new parameter set. A model that produces a lower $\chi^2$ than the previous model is always accepted. A model that produces a higher $\chi^2$ is accepted with probability $\exp(-\Delta\chi^2/2)$. This maps out a probability distribution which can be used to estimate best-fit values and uncertainties for each parameter. More details on the MCMC procedure used in this work can be found in \citet{gibson_2008,gibson_2010}.

The light curve was fitted for $D$ and the central transit time $T_0$. We did not fit the transit width W by allowing it to vary as a free parameter, but to account for uncertainties that may propagate to $D$ and $T_0$, we allowed the transit width to vary within a Gaussian prior by adding a term to the $\chi^2$ statistic so that
\[
\chi^2=\sum_{j=1}^{N}\frac{(f_{{\rm obs},j} - f_{{\rm calc},j})^2}{\sigma^2_j} + 
\frac{(W - 1)^2}{{\sigma_W}^2},
\]
where $\sigma_W$ was set to the fractional error in the transit duration from \citet{hebb_2010}. This led to no significant changes to the parameters and uncertainties determined, but nonetheless we used it in our final analysis for completeness. We allowed the stellar flux to vary either linearly or quadratically as a function of time to normalise the light curves, using a further 2 or 3 normalisation parameters. In order to account for possible correlations between these normalisation parameters and the transit parameters, the normalisation parameters were allowed to vary freely during the fitting process.

An initial MCMC analysis of length 20\,000 was used to estimate the jump functions for $D$, $T_0$, $W$ and the normalisation parameters. The photometric errors $\sigma_j$ were re-scaled so that the best-fitting model had a reduced $\chi^2$ equal to one, as there are often further sources of noise not taken into account from photon statistics alone. This required the errors to be re-scaled by factors of 6.17 and 6.14 when using linear and quadratic normalisation functions, respectively, showing the quadratic function provides a marginally better fit. Such large factors are likely due to variations in pixel-to-pixel sensitivity, which are exaggerated by the dithering pattern required to obtain accurate sky estimates for infrared photometry.

We also examined the residuals for evidence of time-correlated noise using the method of \citet{winn_holman_2008}, where a factor $\beta(\ge1)$ is calculated by analysing the residuals from the best-fit model. The noise should drop by $\sim1/\sqrt{N}$ if it is uncorrelated in time, when the data is averaged into bins of width $N$. However, the noise is often larger by a factor $\beta$ indicating an extra time-correlated source of noise. See \citet{gibson_2008} for a more detailed description of this procedure. We found relatively low levels of time-correlated noise, with the \emph{maximum} value of $\beta\simeq1.08$ for both the linear and quadratic fits, when the residuals were binned into $\sim12$ minute intervals. This indicates that the noise is probably dominated by systematic errors in the randomised dithering pattern that aren't correlated in time. The photometric errors $\sigma_j$ were then rescaled by $\beta$ prior to the final MCMC analysis.

The full MCMC analysis was then performed consisting of five separate chains each of length 100\,000. The first 20\% of each chain were eliminated to keep the initial conditions from influencing the results. The adopted value of each parameter was taken to be the modal value of the probability distribution, and the uncertainties taken as the limits which encompass the central 68.2\% of the probability distribution. To check that the chains had all converged to the same probability distribution, the Gelman \& Rubin Statistic \citep{gelman_rubin_1992} was calculated for each of the free parameters and found to be less than 0.5\% from unity for each, a good sign of mixing and convergence.

As using the quadratic normalisation function only produces a marginally better fit to the data, it is useful to test whether we are `over-fitting' the data. To check which model to use, we evaluated the Bayesian Information Criterion \citep[BIC,][]{schwarz} for the best-fit model using both the linear and quadratic normalisation functions. The BIC is given by
\[
{\rm BIC}_{\chi^2} = \chi^2 + k \ln (n),
\]
where $\chi^2$ is calculated for the best-fit model, $k$ is the number of parameters in the model fit, and $n$ is the number of data points. Increasing the number of free parameters always gives a better fit, but the BIC penalises for higher numbers of model parameters to test the model is not over-fitting the data, and the model that gives the lowest BIC value is the preferred model. Calculating this for the both the linear and quadratic normalisation, indicates that the quadratic normalisation does not improve the fit sufficiently to justify its use, and the results from the model using the linear normalisation function are the adopted measurements. This is true when $\chi^2$ is calculated using the photometric error bars both before and after they are re-scaled. A plot of the normalised light curve and resulting best-fit model using the linear normalisation function is shown in Fig.~\ref{fig:lcv_norm}.

\section[]{Results}
\label{sect:results}

Results from the light curve fits are shown in Table~\ref{tab:res}. The transit depth is found to be 0.366$\pm$0.072\%. As a secondary check on the transit depth, the depth was found by combining the in-transit and out-of-transit data points of the normalised light curve and calculating the depth from the difference. This gave $D$ = 0.350$\pm$0.065\%, in agreement with the depth derived from the MCMC analysis, but with smaller uncertainty as this doesn't take into account uncertainties in the central transit time, transit width, and normalisation parameters.

The central transit time was determined to be 2454921.66790$\pm$0.00190 HJD. Using the ephemeris of \citet{hebb_2010}, this corresponds to a phase of 0.50114$\pm$0.00241, showing that the central eclipse time is consistent with a circular orbit.

\begin{table}
\caption{Parameters and 1$\sigma$ uncertainties as derived from the MCMC fits, and the calculated brightness temperature.}
\label{tab:res}
\begin{tabular}{lccc}
\hline
Parameter & Symbol & Value & Unit \\
\hline
\noalign{\smallskip}
\noalign{\smallskip}
Transit depth &$D$ & 0.366$\pm$0.072 & \%\\
Transit centre& $T_0$ & 2454921.66790$\pm$0.00190 & HJD\\
Brightness temp& $T_B$ & 2540$\pm$180 & K\\
\noalign{\smallskip}
\hline
\end{tabular}
\end{table}

\section[]{Summary and discussion}
\label{sect:summary}

We have detected an occultation of the extrasolar planet system WASP-19 in the NB2090 filter centred at 2.095\,$\umu$m using the HAWK-I instrument. The eclipse depth was measured to be 0.366$\pm$0.072\%, with the transit centre occurring at phase 0.50114$\pm$0.00241, consistent with a circular orbit to well within 1$\sigma$. \citet{anderson_2010}, however, report evidence for a non-zero eccentricity at the 2.6$\sigma$ level. The methods used by \citet{anderson_2010} to measure the phase of the eclipse, and to correct for systematics, particularly the trends in the light curve which can considerably effect transit times \citep[e.g.][]{gibson_2009}, differs from that adopted in our analysis, and this difference is the most likely source of the apparent discrepancy.

The brightness temperature may be calculated from the depth, taking the values and uncertainties for the stellar temperature and planet-to-star radius ratio from \citet{hebb_2010}. This results in a brightness temperature of 2540$\pm$180\,K, considerably larger than the (zero-albedo) equilibrium temperature given in \citet{hebb_2010} of 2009$\pm$26\,K. \citet{anderson_2010} report an eclipse depth of $0.259^{+0.046}_{-0.044}\%$ in the H-band, corresponding to a brightness temperature of 2560$\pm$130\,K. This is again considerably higher than the equilibrium temperature, and also the equilibrium temperature given no redistribution of heat to the nightside, which they calculate as $\sim2400$\,K. Our 2.095\,$\umu$m measurement is consistent with this temperature.

This indicates that there is poor redistribution of heat to the night side of the planet, consistent with the pM class of planets in the classification of \citet{fortney_2008}. They also conclude that planets hot enough to have significant TiO and VO absorption in their upper atmospheres would show temperature inversions, and may appear anomalously bright in the infrared due to molecular emission bands, which may help explain the temperature excesses here. Other ground based K-band secondary eclipse measurements of very hot Jupiters reached similar conclusions \citep{demooij_snellen,gillon_2009,alonso_2010}. Further observations at infrared wavelengths are required to confirm temperature inversions and possibly measure molecular emission for WASP-19, which should prove an interesting target for future studies.

\section*{Acknowledgments}

We would like to thank the VLT support staff for advice during the preparation and execution of the observing run, particularly M. Petr-Gotzens. We also thank the referee, D. Deming, for comments which improved the clarity of this manuscript.

%\bibliography{MNRAS_WASP19} % your references xx.bib 
%\bibliographystyle{aa} % style xx.bst

\label{lastpage}

\end{document}